\documentclass[11pt, oneside]{article}
\usepackage[a4paper, left=1.0in, right=1.0in, top=1.0in, bottom=1.5in, bindingoffset=0mm]{geometry}

\usepackage{newtxtext}
\usepackage{newtxmath}
\usepackage[onehalfspacing]{setspace}
\usepackage[parfill]{parskip}
\usepackage[version=4]{mhchem}
\usepackage{siunitx}
\usepackage{amsmath}
\usepackage{booktabs}
\usepackage{xcolor}

\usepackage{graphicx}
\graphicspath{{fig/}}

\usepackage[
    backend=biber, 
    style=authoryear,
    sorting=nyt, 
    url=false, 
    uniquelist=false, 
    hyperref=auto, 
    giveninits=true, 
    maxnames=3, 
    minnames=1, 
]{biblatex}
\usepackage{hyperref}
\hypersetup{
    colorlinks=true,  
    linkcolor=blue,   
    citecolor=blue,   
    urlcolor=blue     
}
\usepackage{url}
\newcommand{\CSR}{\mathrm{CSR}}

\newcommand{\Eimage}{E_{\mathrm{image}}}
\newcommand{\zc}{z_{\mathrm{c}}}
\newcommand{\Pt}{P_{\mathrm{t}}}
\newcommand{\Pn}{P_n}

\title{Estimation of the electric field in atom probe tomography experiments using charge state ratios}
\author{Levi Tegg$^{1,2}$, Leigh T. Stephenson$^{2}$ and Julie M. Cairney$^{1,2*}$}
\date{\small{
  $^1$ School of Aerospace, Mechanical and Mechatronic Engineering, The University of Sydney, Camperdown, NSW 2006, Australia\\
  $^2$ Australian Centre for Microscopy and Microanalysis, The University of Sydney, Camperdown, NSW 2006, Australia\\
  $^{*}$ Corresponding author:  \href{mailto:Julie.Cairney@sydney.edu.au}{Julie.Cairney@sydney.edu.au}
}}

\begin{document}

\maketitle

\section*{Abstract}
\autocite{Kingham_1982} provided equations for the probability of observing higher charge states in atom probe tomography (APT) experiments.
These ``Kingham curves'' have wide application in APT, but cannot be analytically transformed to provide the electric field in terms of the easily-measured charge state ratio (CSR).
Here we provide a numerical scheme for the calculation of Kingham curves and the variation in electric field with CSR.
We find the variation in electric field with CSR is well-described by a simple two or three-parameter equation, and the model is accurate to most elements and charge states.
The model is applied to experimental APT data of pure aluminium and a microalloyed steel, demonstrating that the methods described in this work can be easily applied to a variety of APT problems to understand electric field variations.

\section{Introduction}
In atom probe tomography (APT), atoms at the apex of a needle-shaped specimen are both field-evaporated and accelerated towards a position-sensitive detector using strong electric fields. 
The rate of field evaporation is controlled by varying the applied voltage between 0.5~kV and 10~kV, and with brief pulses of energy provided by a laser or variations in the voltage.
By measuring the time between the pulse and the ion reaching the detector, a spectrum of mass-to-charge ratios (or ``mass spectrum'') can be produced \autocite{GaultTextbook_2012}.
Although only a single ionisation (i.e., atom A to ion \ce{A+}) is required for field-evaporation to occur, higher charge states (such as \ce{A++} and \ce{A+++}) are commonly observed in mass spectra.

\autocite{Kingham_1982} provided a model for why higher charge states are observed using post-evaporation ionisation (or ``post ionisation'').
In that work, expressions were derived for the probability of an electron tunnelling from the ion to the surface, $\Pt$, in terms of the local electric field at the apex of the specimen,~$F$.
Manipulating $\Pt$ provides the probability $\Pn$ of observing a charge state $n$, and \autocite{Kingham_1982} plots curves of $\Pn(F)$ for 24 elements.
The importance of these ``Kingham curves'' to the atom probe community is indicated by their reproduction in APT textbooks \autocite{Miller_1996, GaultTextbook_2012, Miller_Forbes_2014} and reference in several hundred journal articles.
Knowledge of the local electric field is valuable in studies of hydrogen in steels and alloys \autocite{Felfer_2012,Chang_2018, Breen_2020}, carbides \autocite{Thuvander_2011}, semiconductors \autocite{Mancini_2014, Hans_Schneider_2020, Cuduvally_2022}, and into the fundamentals of field evaporation and atom probe instrumentation \autocite{Lam_Needs_1992, Vella_2021, Breen_2023, Tegg_2023}.
Variations in electric field can result in distortions to microstructural features (known as ``local magnification'') \autocite{LarsonSpatial_2013, Lawitzki_2021}, changes to multiple-ion evaporation \autocite{Saxey_2011, Peng_2018}, and can affect complex ion evaporation \autocite{GaultTextbook_2012, Larson_2013}, so knowledge of the electric field in a specimen can significantly aid in data interpretation.

The curves in \autocite{Kingham_1982} give $\Pn(F)$: the probability of observing a certain charge state based on the electric field.
However, atom probe researchers more commonly seek $F(\Pn)$: the electric field calculated using the measured charge state, or the charge state ratio (CSR).
This is because several phenomena in atom probe data can be explained by variations in local electric field \autocite{Felfer_2012, Chang_2018, Breen_2023}.
The expressions for $\Pt(F)$ cannot be analytically transformed to provide $F(\Pt)$.
This means that atom probe researchers wanting to calculate electric fields must measure their CSR and either read $F$ graphically from published Kingham curves \autocite{Kingham_1982, GaultTextbook_2012, Miller_Forbes_2014}, or calculate $\Pn(F)$ themselves.

Here, we present a numerical method for calculating Kingham curves for 60 elements. 
We also calculate $F(\CSR)$, and find that a simple approximate expression describes for $F(\CSR)$ across $10^{-4} \leq \CSR \leq 10^{4}$ for each element and combination of charge states up to $0 \leq F \leq 70$~V/nm. 
This expression can be used to quickly estimate the electric field to within <1\% of the values predicted from Kingham curves , and is easily applied to experimental data.
We provide graphs of $F(\CSR)$, a table of coefficients, and the Python 3.11 code used to calculate $\Pn(F)$ and $F(\CSR)$.
Ethis work serves as a pratical starting point for estimating surface electric field in order to better understand various phneomna in experimental atom probe data.

\section{Theory and implementation}

\subsection{Kingham's model} \label{section:kingham}

Here we will partially reproduce the model of \autocite{Kingham_1982} in order to introduce our computational method.
We also include some of the modifications or clarifications made by \autocite{Andren_1984, Lam_Needs_1992, Cuduvally_2022}.
Expressions are introduced in the order they are calculated in the associated Python code, and some have been re-arranged from that shown in the source material.
As in \autocite{Kingham_1982}, the expressions below are given in Hartree atomic units (au).
The reduced Planck constant, the mass of the electron, and the electronic charge are all $\hbar = m_{\mathrm{e}} = e = 1$, respectively.
Energy units are 1 $E_{\mathrm{h}}$ = 27.2 eV, distance units are Bohr radii 1 $a_0$ = 0.053 nm, and electric field units are 1 $E_{\mathrm{h}}/ea_0$ = 514.2 V/nm.

\begin{figure}[t!]
  \centering
  \includegraphics[width=3.7in]{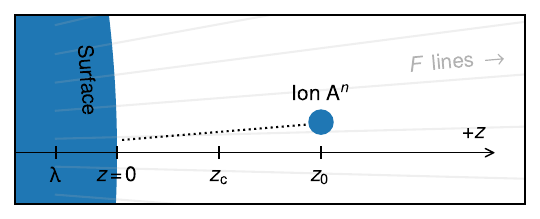}
  \caption{A schematic of the model in \autocite{Kingham_1982}. An ion \ce{A^{n}} in charge state $n \geq 1$ is field-evaporated and travels along (or very close to) electric field lines (``$F$ lines $\rightarrow$'') at velocity $u(z_0)$. Post-ionisation to charge state $n + 1$ is possible at ion-tip distances $z_0$ that are greater than a critical distance, $z_\mathrm{c}$. The electric field $F$ penetrates up to a distance $\lambda$ below the tip surface at $z = 0$.}
  \label{fig:schematic}
\end{figure}

In \autocite{Kingham_1982}'s model, post-ionisation occurs when an electron tunnels from the field-evaporated ion \ce{A^{n}} to an unoccupied electronic state on the surface of the tip.
The ion initially evaporates in charge state $n_\mathrm{i}$ but is observed at the detector with charge state $n \geq n_\mathrm{i}$.
The probability of tunnelling $\Pt$ from a given charge state $n$ to $n + 1$ depends on the distance between the tip surface and the ion ($z_0$) and the ionisation energy of the the $n+1$ state ($I_{n+1}$, hereafter $I$).
Figure \ref{fig:schematic} shows a schematic of the model system.

Tunnelling becomes likely at a critical distance, $\zc$, where the energy of the least-tightly bound electron on \ce{A^n} is of a greater energy than the Fermi level at the tip surface.
The ionisation energy of the outermost electron of the ion is less than that of same ion would be in a field-free environment.
\autocite{Cuduvally_2022} calculate this energy as
\begin{equation} \label{eq:Fzc}
  F\zc = I - \phi_0 - \Delta \Eimage - \lambda F \,.
\end{equation}
Here, $\phi_0$ is the zero-field work function, $\lambda = 0.8$ is the distance the field penetrates into the tip surface in $a_0$ units \autocite{Lang_Kohn_1973}, and $\Eimage$ is the image potential, given by
\begin{equation} \label{eq:Eimage}
  \Eimage = \frac{n^2}{4z_c}
\end{equation}
at $\zc$.
Here we follow \autocite{Cuduvally_2022} by assuming the Stark shift is negligible.
Equation \ref{eq:Fzc} is a quadratic equation in $\zc$, and the real solution with the positive root is
\begin{equation} \label{eq:zc}
  \zc(n) = \begin{cases}
    \dfrac{(I - \phi_0) + \sqrt{(I - \phi_0)^2 - (2n+1)F}}{2F} & \text{for } (I - \phi_0)^2 \geq (2n+1)F\\
    0 & \text{elsewhere}
  \end{cases} \\.  
\end{equation}

The rate constant of post-ionisation, $R$, is found from the electron probability flux of the outermost electron orbital through a surface $S$ perpendicular to the field direction,
\begin{equation} \label{eq:R1}
  R = \int_S \left| \psi (r, \theta) ^2 \right| v_z \mathrm{d} S \,,
\end{equation}
where $\psi(r, \theta)$ is an s-type electron wavefunction in polar co-ordinates, and $v_z$ is the electron velocity normal to the metal surface.
\autocite{Kingham_1982} provides the approximate solution when the angle between the ion trajectory and the field, $\theta$, is small:
\begin{equation} \label{eq:R3}
  R(z_0) = \frac{6 \pi A^2 \nu F}{2^{5/2} ( I^{3/2} - G ) } \left( \frac{16 I^2}{ZF} \right)^{Z \sqrt{\frac{2}{I}}}
  \mathrm{exp} \left[ \frac{-2^{5/2} (I^{3/2} - G )  }{3F} + \frac{Z\sqrt{\frac{2}{I}}}{3} \right] \\.
\end{equation}
Here, $\mathrm{exp}\left[ x \right] = \mathrm{e}^{x}$ is the exponential function, $A^2\nu$ is given by
\begin{equation} \label{eq:A2nu}
  A^2\nu = \frac{I}{6 \pi m \ \mathrm{exp} \left[ \frac{2}{3} \right]} \,,
\end{equation}
and $G$ is introduced here to simplify the typesetting of equation \ref{eq:R3}:
\begin{equation} \label{eq:R3helper}
  G(z_0) = \begin{cases}
    \left( I - \dfrac{ZF}{I} - F z_0 \right)^{3/2} & \text{for }z_0 \leq \frac{I}{F} - \frac{Z}{I}\\
    0 & \text{elsewhere}
  \end{cases} \\.
\end{equation}
$Z$ is found by \autocite{Kingham_1982} by fitting the calculated $\Pt$ to experimental values for post-ionisation from \ce{Rh+} to \ce{Rh++} from \autocite{Ernst_1979}:
\begin{equation} \label{eq:Z}
  Z(z_0) = n + 1 + \frac{4.5}{z_0} \\.
\end{equation}

Lastly, the ion velocity $u$ is found from
\begin{equation} \label{eq:u}
  \frac{1}{2}m_{\mathrm{ion}} u(z_0)^2  = (z_0 + \lambda)nF - \sum_{r = n_\mathrm{i}}^{n-1} \left( ( \zc(r) + \lambda)F  + \frac{2 r + 1}{4(\zc(r)+\lambda)} \right) + 
  \frac{n^2}{4(z_0 + \lambda)} + \sqrt{n_\mathrm{i}^3 F} \\,
\end{equation}
where $m_{\mathrm{ion}}$ is the mass of the ion \autocite{Kingham_1982, Andren_1984}.
Equations \ref{eq:R3} and \ref{eq:u} are used to calculate the probability of post-ionisation at each field and charge state,
\begin{equation} \label{eq:Pt}
  \Pt(n, F) = 1 - \mathrm{exp} \left[ - \int_{z_\mathrm{c}}^\infty \frac{R(z_0, n, F)}{u(z_0, n, F)} \mathrm{d}z_0 \right] \\.
\end{equation}
These probabilities are more commonly expressed as the probability of observing a certain change state at the detector, $\Pn$, which is found from
\begin{equation} \label{eq:Pn1}
  P_{n=1} = 1 - \Pt(n = 2)
\end{equation}
for the $n = 1$ (+) charge state, and
\begin{equation} \label{eq:Pnn}
  \Pn = P_{\mathrm{t}}(n) - P_{\mathrm{t}}(n+1)
\end{equation}
for the $n \geq 2$ (++, +++, and ++++) charge states.
The CSR can be found from equations \ref{eq:Pn1} and \ref{eq:Pnn}, or from an experimental mass spectrum, by
\begin{equation} \label{eq:CSRdefinition}
  \CSR = \frac{P_{n+1}}{P_{n}} = \frac{C_{i, n+1}}{C_{i, n}} \\,
\end{equation}
where $C$ is the ranged counts in the higher ($n+1$) or lower ($n$) charge state for a given element or isotope~($i$).
Alternatively, $C$ is the ionic concentration of a given $i$ at $n$ in reconstructed APT data, providing each~$i$ and~$n$ are sampled from the same reconstructed volume.

\subsection{Numerical implementation} \label{section:implementation}
In this work, $\Pt$ is calculated using the equations given in the previous section.
The main calculation loop is performed in the Hartree atomic units but the input parameters and the output plots are expressed in eV, nm and V/nm.
Ionisation energies are taken from \autocite{NIST_ASD} and work functions from \autocite{Michaelson_1977}.
The work functions used are those for polycrystalline solids in their room temperature phases, and work function for C is that of polycrystalline graphite.
Electric fields are calculated across $0.01 \leq F \leq 100$~V/nm in 1000 steps with equal spacing, charge states $1 \leq n \leq 4$, and distances $10^{-2} \leq z_0 \leq 10^{6}$~nm in 5000 steps with exponentially-increasing spacing.
The main calculation loop consists of a loop over elements, then electric fields, then charge states, with $\Pt$ calculated for each.
$P_n$ and $\CSR$ curves are then calculated for each element.

\autocite{Kingham_1982} does not provide equation \ref{eq:zc}, and \autocite{Cuduvally_2022} does not indicate which root to use, or how to handle a complex solution.
We find agreement with their $\Pn(F)$ curves if we use the positive root and set $\zc = 0$ when a complex solution would be obtained, as this would represent a $\zc$ inside the metal.
Once $\zc(n, F)$ has been found, subsequent calculations involving $z_0$ are only performed for the domain $z_0 > \zc$.
Like \autocite{Cuduvally_2022}, we applied the correction to the exponent in equation \ref{eq:R3} given by \autocite{Lam_Needs_1992}, but do not use their expressions for $R$.
Equation \ref{eq:u} is not fully defined in \autocite{Kingham_1982}, and not expressed for $n > 1$ in \autocite{Cuduvally_2022}. 
\autocite{Andren_1984} clarifies that the two sums presented in equation 3.39 of \autocite{Kingham_1982} should be combined into one sum, as shown here in equation \ref{eq:u}.
We have assumed the initial charge state of the field-evaporated ion is always $n_\mathrm{i} = 1$.
Like \autocite{Kingham_1982}, we do not observe significant differences in the results for $n_\mathrm{i} \geq 2$.
The integral in equation \ref{eq:Pt} is calculated using the trapezoidal method, and thus the result is sensitive to the array of $z_0$ used. 
We use an exponentially-spaced array to ensure accurate $\zc$ calculation at low $z$, while ensuring sufficiently high $z_0$ are included to suit the infinite upper limit.

The Python code used to perform these calculations is provided as supplementary information to this manuscript.
The code was written for Python 3.11.5 with \texttt{numpy} version 1.19.2 \autocite{numpy_2020}, \texttt{scipy} version 1.5.2 \autocite{SciPy_2020}, and \texttt{matplotlib} version 3.3.2 \autocite{matplotlib_2007}.
Key equations from section \ref{section:kingham} are expressed as functions, allowing for a modular approach to the calculations of $\Pt$ and $\CSR$.
Atomic and material parameters are also supplied in a comma-separated variable file.
The same code was used to prepare the figures shown in this manuscript.

\subsection{Expression for the electric field} \label{section:approximation}
Plots of $F(\CSR)$ were produced from the specified arrays of $F$ and the calculated arrays of $\CSR$.
We found that equation \ref{eq:FCSR} described plots of $F(\mathrm{CSR})$ with reasonable accuracy:
\begin{equation} \label{eq:FCSR}
  F = a \left(1 - \frac{b}{\CSR^{0.3} + b + 0.256} \ \right)\ \   \mathrm{V/nm} \\.
\end{equation}
The coefficients $a$ and $b$ were found for each element and $\CSR$ using the \texttt{curve\_fit} function from the \texttt{scipy.optimize} library \autocite{SciPy_2020}.
The fit was performed over $10^{-4} \leq \CSR \leq 10^{4}$ as it was felt this is the maximum range where CSR can be accurately calculated in typical atom probe data.
This assumption is covered in the discussion section.

\subsection{Application to experiment} \label{section:methods-experiment}
Equation \ref{eq:FCSR} was used to estimate electric field using CSRs in experimental atom probe data.
Data from a pure Al specimen was collected using a Cameca Invizo 6000 \autocite{Tegg_2023} by using voltage-pulsed acquisition with 20\% pulse fraction, 200 kHz pulse rate, 2\% target detection rate and at 50 K temperature.
Data from a microalloyed martensitic steel \autocite{Lin_2020} was collected using a Cameca Invizo 6000 by using laser-pulsed acquisition with 400 pJ laser pulse energy, 200 kHz pulse rate, 4\% target detection rate and at 50~K temperature.
Data reconstruction and analysis was performed using the IVAS module within AP Suite 6.3.
The Al dataset was reconstructed in detector space to allow for easy calculation of isotope-specific field evaporation images.
The microalloyed steel was reconstructed by calculating the sample radius from the standing voltage and the evaporation field of Fe \autocite{Tsong_1978}. 
The image compression factor and field factor determined by inspection of crystallographic poles \autocite{GaultTextbook_2012}.
Further details of the analysis methods are provided in section \ref{section:results-application}.

\section{Results}

\subsection{Calculated fields and CSRs}  \label{section:results-calculation}

\begin{figure}[t!]
  \centering
  \includegraphics[width=3.46in]{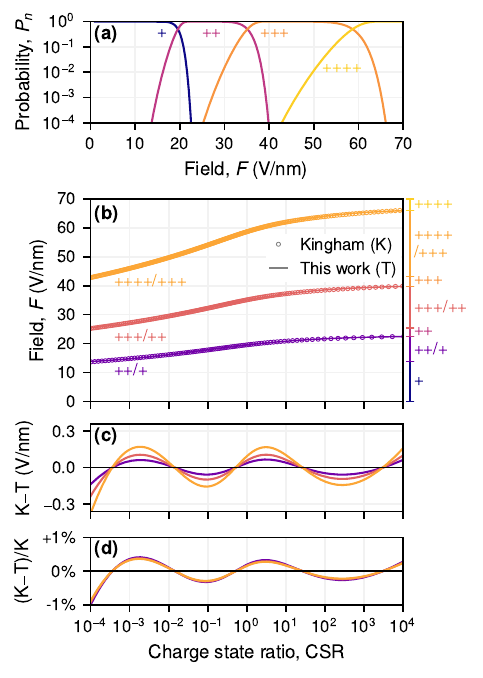}
  \caption{Illustration of the method reported here.
  (a) Kingham curves for W up to ++++, showing the probability ($\Pn$) of observing each charge state given the local electric field ($F$). (b) Ratio of consecutive pairs of Kingham curves expressed in terms of the charge state ratio (CSR). Curves calculated from (a) shown as hollow circles (K), curves fit with equation \ref{eq:FCSR} shown as solid lines (T). Annotations on the right vertical axis indicate the ranges for each charge state in $F$. (c) The difference ($\mathrm{K} - \mathrm{T}$) between the Kingham curve and the model in this work. (d) The difference from (c) expressed as a percentage error (i.e. $100 \times $($\mathrm{K} - \mathrm{T}$)/$\mathrm{K}$) of the Kingham curves. (b-d) share a horizontal axis.}
  \label{fig:one-element}
\end{figure}

Figure \ref{fig:one-element} illustrates the method described in sections \ref{section:kingham}--\ref{section:approximation}. 
Figure \ref{fig:one-element}(a) shows Kingham curves for W between $0 \leq F \leq 70$~V/nm, $10^{-4} \leq P_n \leq 10^{4}$, and for charge states + to ++++.
As the field increases, the probability of observing the + charge state falls, and the probabilities of observing higher charge state rise and fall successively.
Figure \ref{fig:one-element}(b) shows the calculation of $F(\CSR)$ for the ++/+, +++/++, and ++++/+++ CSRs.
Data points calculated by dividing successive pairs of Kingham curves in (a) are shown in (b) as hollow circles. 
For the purposes of error estimation, these data are labelled ``K''.
Fits of K using equation \ref{eq:FCSR} are shown in figure \ref{fig:one-element}(b) as solid lines.
For error estimation these curves are labelled ``T''.
Agreement between K and T is excellent: figure \ref{fig:one-element}(c) shows the difference between the $F(\CSR)$ calculated from Kingham curves and the model using equation \ref{eq:FCSR}, i.e. the residual $\mathrm{K} - \mathrm{T}$.
The modelled $F(\CSR)$ generally stays within $\approx 0.2$~V/nm of the values calculated from Kingham curves.
The oscillation in the residual indicates that higher-order polynomial terms could be used to better model~K.
However, figure \ref{fig:one-element}(d) shows the residual relative to K to provide the percentage error between K~and~T.
Across $10^{-3} \leq \mathrm{CSR} \leq 10^3$ the percentage error is <0.5\%, and only increases to $\approx 1.0$\% below CSR~$< 10^{-3}$.
As such, we felt there was no need to include additional free parameters to equation~\ref{eq:FCSR}.
Note that this residual and percentage error refers to the difference between the $F(\CSR)$ calculated from Kingham curves and from the model in equation \ref{eq:FCSR}.
Other sources of uncertainty are not quantified here and are covered in the discussion section.
The annotations on the right vertical axis of (b) indicate the charge states expected for different ranges in $F$.
The ranges where a single charge state is observed are relatively narrow, for example the +++ state is expected only between $\approx 40$~V/nm and 43~V/nm.
This idea is also discussed further in later sections.

\begin{figure}[t!]
  \centering
  \makebox[\textwidth][c]{\includegraphics[width=7.0in]{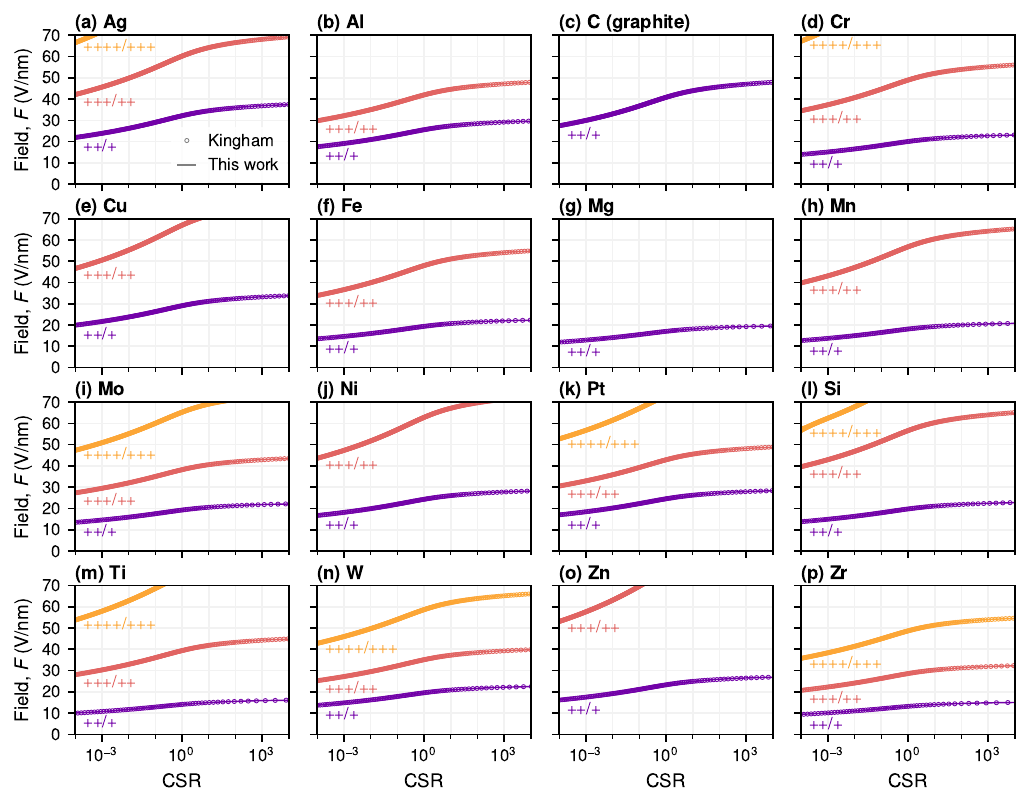}}
  \caption{$F(\mathrm{CSR})$ curves for 16 elements, with CSR up to ++++/+++ and $F \leq 70$~V/nm. As with figure~\ref{fig:one-element}, hollow circles show data calculated from successive Kingham curves, solid lines show models fit with equation \ref{eq:FCSR}. Fitting coefficients are shown in table 1 and figure \ref{fig:model-params}}
  \label{fig:many-elements}
\end{figure}

\begin{table}[b!]
  \centering
  \footnotesize
  \caption{Model parameters $a$ and $b$ from equation \ref{eq:FCSR} for the elements and CSRs explored in this study.}
  \label{table:all-parameters}
  \begin{tabular}[t]{c c c}
  \begin{tabular}[t]{llll}
    \toprule
     &     & $a$ & $b$ \\ \midrule
     Ag & ++/+ & 37.89 & 0.2261 \\
     Ag & +++/++ & 69.93 & 0.2053 \\
     Ag & ++++/+++ & 107.4 & 0.1915 \\
     Al & ++/+ & 29.99 & 0.2196 \\
     Al & +++/++ & 48.36 & 0.1939 \\
     As & ++/+ & 28.88 & 0.2153 \\
     As & +++/++ & 47.45 & 0.1910 \\
     Au & ++/+ & 33.55 & 0.2199 \\
     Au & +++/++ & 52.56 & 0.1938 \\
     Au & ++++/+++ & 90.73 & 0.1834 \\
     B & ++/+ & 51.54 & 0.2417 \\
     Ba & ++/+ & 9.067 & 0.1755 \\
     Ba & +++/++ & 74.19 & 0.2083 \\
     Bi & ++/+ & 23.51 & 0.2065 \\
     Bi & +++/++ & 38.83 & 0.1831 \\
     Bi & ++++/+++ & 92.05 & 0.1840 \\
     C & ++/+ & 48.48 & 0.2384 \\
     Ca & ++/+ & 12.68 & 0.1889 \\
     Cd & ++/+ & 24.24 & 0.2091 \\
     Cd & +++/++ & 80.21 & 0.2102 \\
     Cd & ++++/+++ & 115.6 & 0.1943 \\
     Ce & ++/+ & 10.76 & 0.1812 \\
     Ce & +++/++ & 25.08 & 0.1701 \\
     Ce & ++++/+++ & 62.82 & 0.1725 \\
     Co & ++/+ & 24.85 & 0.2112 \\
     Co & +++/++ & 65.41 & 0.2038 \\
     Cr & ++/+ & 23.34 & 0.2098 \\
     Cr & +++/++ & 56.59 & 0.1991 \\
     Cr & ++++/+++ & 108.9 & 0.1933 \\
     Cs & ++/+ & 43.76 & 0.2328 \\
     Cs & +++/++ & 64.34 & 0.2042 \\
     Cu & ++/+ & 34.17 & 0.2229 \\
     Cu & +++/++ & 78.01 & 0.2099 \\
     Eu & ++/+ & 11.28 & 0.1825 \\
     Eu & +++/++ & 37.07 & 0.1829 \\
     Eu & ++++/+++ & 83.60 & 0.1818 \\
     Fe & ++/+ & 22.54 & 0.2080 \\
     Fe & +++/++ & 55.42 & 0.1980 \\
     Ga & ++/+ & 34.77 & 0.2228 \\
     Ga & +++/++ & 55.67 & 0.1971 \\
     Gd & ++/+ & 12.90 & 0.1867 \\
     Gd & +++/++ & 25.82 & 0.1706 \\
     Gd & ++++/+++ & 89.13 & 0.1838 \\
     Ge & ++/+ & 21.68 & 0.2051 \\
     Ge & +++/++ & 67.06 & 0.2036 \\
     Ge & ++++/+++ & 94.12 & 0.1886 \\
     Hf & ++/+ & 18.38 & 0.1985 \\
     Hf & +++/++ & 30.74 & 0.1758 \\
     Hf & ++++/+++ & 51.56 & 0.1652 \\
     Hg & ++/+ & 29.21 & 0.2147 \\
     Hg & +++/++ & 68.25 & 0.2030 \\
     Hg & ++++/+++ & 104.8 & 0.1889 \\
     \bottomrule
  \end{tabular} 
  &
  \begin{tabular}[t]{llll}
    \toprule
     &     & $a$ & $b$ \\ \midrule
     In & ++/+ & 29.72 & 0.2166 \\
     In & +++/++ & 46.52 & 0.1904 \\
     In & ++++/+++ & 136.5 & 0.2038 \\
     Ir & ++/+ & 24.35 & 0.2082 \\
     Ir & +++/++ & 46.19 & 0.1894 \\
     Ir & ++++/+++ & 72.63 & 0.1760 \\
     K & ++/+ & 78.87 & 0.2599 \\
     La & ++/+ & 11.19 & 0.1824 \\
     La & +++/++ & 22.73 & 0.1669 \\
     Lu & ++/+ & 17.28 & 0.1966 \\
     Lu & +++/++ & 26.76 & 0.1713 \\
     Lu & ++++/+++ & 91.97 & 0.1844 \\
     Mg & ++/+ & 19.75 & 0.2045 \\
     Mn & ++/+ & 21.01 & 0.2046 \\
     Mn & +++/++ & 65.84 & 0.2035 \\
     Mo & ++/+ & 22.37 & 0.2071 \\
     Mo & +++/++ & 43.91 & 0.1893 \\
     Mo & ++++/+++ & 74.40 & 0.1782 \\
     Nb & ++/+ & 17.85 & 0.1992 \\
     Nb & +++/++ & 37.76 & 0.1841 \\
     Nb & ++++/+++ & 65.21 & 0.1739 \\
     Nd & ++/+ & 10.44 & 0.1802 \\
     Nd & +++/++ & 29.72 & 0.1757 \\
     Nd & ++++/+++ & 75.27 & 0.1784 \\
     Ni & ++/+ & 28.52 & 0.2209 \\
     Ni & +++/++ & 73.31 & 0.2123 \\
     Os & ++/+ & 24.36 & 0.2083 \\
     Os & +++/++ & 37.30 & 0.1821 \\
     Os & ++++/+++ & 76.13 & 0.1776 \\
     Pb & ++/+ & 19.31 & 0.1996 \\
     Pb & +++/++ & 59.03 & 0.1977 \\
     Pb & ++++/+++ & 81.00 & 0.1851 \\
     Pd & ++/+ & 31.45 & 0.2190 \\
     Pd & +++/++ & 63.07 & 0.2016 \\
     Pd & ++++/+++ & 95.21 & 0.1863 \\
     Pt & ++/+ & 28.66 & 0.2141 \\
     Pt & +++/++ & 49.33 & 0.1917 \\
     Pt & ++++/+++ & 83.27 & 0.1805 \\
     Rb & ++/+ & 59.42 & 0.2455 \\
     Rb & +++/++ & 88.28 & 0.2168 \\
     Re & ++/+ & 23.31 & 0.2068 \\
     Re & +++/++ & 43.16 & 0.1873 \\
     Re & ++++/+++ & 69.62 & 0.1747 \\
     Rh & ++/+ & 27.53 & 0.2143 \\
     Rh & +++/++ & 56.56 & 0.1979 \\
     Rh & ++++/+++ & 80.27 & 0.1817 \\
     Ru & ++/+ & 23.91 & 0.2092 \\
     Ru & +++/++ & 48.03 & 0.1922 \\
     Ru & ++++/+++ & 91.46 & 0.1851 \\
     Sb & ++/+ & 23.43 & 0.2074 \\
     Sb & +++/++ & 38.30 & 0.1834 \\
     Sb & ++++/+++ & 86.49 & 0.1823 \\

    \bottomrule
  \end{tabular}
  &
  \begin{tabular}[t]{llll}
    \toprule
     &     & $a$ & $b$ \\ \midrule
     Sc & ++/+ & 14.57 & 0.1933 \\
     Sc & +++/++ & 37.24 & 0.1850 \\
     Se & ++/+ & 36.79 & 0.2241 \\
     Se & +++/++ & 58.46 & 0.1981 \\
     Se & ++++/+++ & 83.33 & 0.1819 \\
     Si & ++/+ & 23.00 & 0.2094 \\
     Si & +++/++ & 65.71 & 0.2049 \\
     Si & ++++/+++ & 92.78 & 0.1893 \\
     Sm & ++/+ & 10.97 & 0.1817 \\
     Sm & +++/++ & 33.52 & 0.1796 \\
     Sm & ++++/+++ & 78.88 & 0.1798 \\
     Sn & ++/+ & 18.46 & 0.1990 \\
     Sn & +++/++ & 54.46 & 0.1958 \\
     Sn & ++++/+++ & 75.58 & 0.1810 \\
     Sr & ++/+ & 10.94 & 0.1823 \\
     Ta & ++/+ & 22.30 & 0.2054 \\
     Ta & +++/++ & 32.17 & 0.1777 \\
     Ta & ++++/+++ & 56.44 & 0.1680 \\
     Tb & ++/+ & 11.78 & 0.1838 \\
     Tb & +++/++ & 28.96 & 0.1743 \\
     Tb & ++++/+++ & 70.68 & 0.1759 \\
     Te & ++/+ & 28.83 & 0.2146 \\
     Te & +++/++ & 45.75 & 0.1894 \\
     Te & ++++/+++ & 64.11 & 0.1733 \\
     Th & ++/+ & 12.92 & 0.1866 \\
     Th & +++/++ & 20.73 & 0.1632 \\
     Th & ++++/+++ & 38.57 & 0.1561 \\
     Ti & ++/+ & 16.24 & 0.1968 \\
     Ti & +++/++ & 45.33 & 0.1916 \\
     Ti & ++++/+++ & 85.58 & 0.1841 \\
     Tl & ++/+ & 34.21 & 0.2205 \\
     Tl & +++/++ & 52.02 & 0.1933 \\
     Tl & ++++/+++ & 115.1 & 0.1899 \\
     U & ++/+ & 11.92 & 0.1838 \\
     U & +++/++ & 24.02 & 0.1679 \\
     U & ++++/+++ & 61.83 & 0.1711 \\
     V & ++/+ & 18.68 & 0.2015 \\
     V & +++/++ & 51.08 & 0.1956 \\
     V & ++++/+++ & 98.78 & 0.1889 \\
     W & ++/+ & 22.72 & 0.2061 \\
     W & +++/++ & 40.21 & 0.1849 \\
     W & ++++/+++ & 66.63 & 0.1733 \\
     Y & ++/+ & 13.26 & 0.1886 \\
     Y & +++/++ & 25.90 & 0.1715 \\
     Zn & ++/+ & 27.21 & 0.2140 \\
     Zn & +++/++ & 89.73 & 0.2150 \\
     Zr & ++/+ & 15.17 & 0.1933 \\
     Zr & +++/++ & 32.62 & 0.1793 \\
     Zr & ++++/+++ & 55.09 & 0.1684 \\

    \bottomrule
  \end{tabular} 
\end{tabular}
  \end{table}

\begin{figure}[t!]
  \centering
  \includegraphics[width=3.46in]{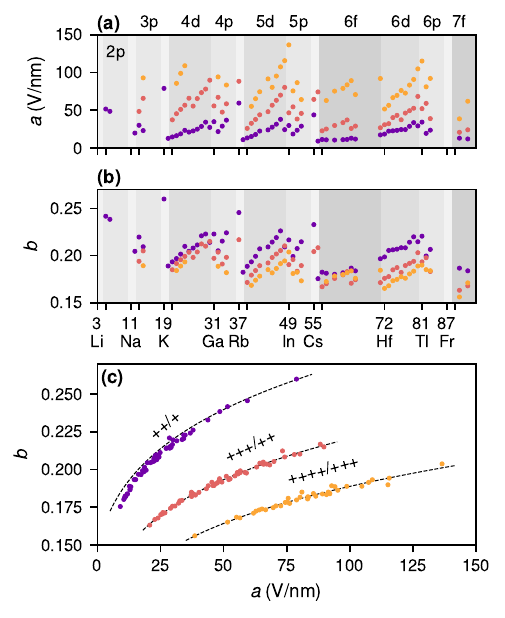}
  \caption{Variation in model parameters (a) $a$ and (b) $b$ with atomic number for the (purple) ++/+, (pink) +++/++ and (orange) ++++/+++ CSRs. Horizontal axis ticks and shading indicate blocks from the periodic table of the elements. (c) Relationship between model parameters $b$ and $a$ for all elements. Dashed lines indicate a fit using equation \ref{eq:abfit}.}
  \label{fig:model-params}
\end{figure}

Figure \ref{fig:many-elements} shows the $F(\CSR)$ curves for 16 elements commonly studied using APT. 
As with figure \ref{fig:one-element}, hollow circles denote $F(\CSR)$ data calculated from successive pairs of Kingham curves, and solid lines denote fits using equation \ref{eq:FCSR}.
The fit parameters are listed in table \ref{table:all-parameters} and plotted in figure \ref{fig:model-params}.
Equation~\ref{eq:FCSR} accurately describes the ++/+ and +++/++ $F(\CSR)$ curves for all elements studied.
There are small variations in the shape of the ++++/+++ $F(\CSR)$ curves at $\CSR \leq 10^{-3}$ for some elements, such as Si shown in figure \ref{fig:many-elements}(l), which make the model less accurate in these regions.

Figure \ref{fig:model-params}(a,b) shows the variation in model parameters $a$ and $b$ with atomic number.
The shading indicates the s, p, d and f blocks of the elements.
With reference to equation \ref{eq:FCSR}, the parameter $a = F(\CSR \rightarrow \infty)$ represents the electric field at infinite CSR and generally increases along a block, e.g., the fourth period d-block transition metals between Ti and Cu.
The parameter $b$ is related to the $F(\CSR \rightarrow 0)$, by
\begin{equation}
  b = 0.256 \left( \frac{F(\CSR \rightarrow \infty)}{F(\CSR \rightarrow 0)} - 1 \right) \\.
\end{equation}
Figure \ref{fig:model-params}(c) shows the relationship between $b$ and $a$. 
It was found that $b$ can be described by $a$ with an equation of the form
\begin{equation} \label{eq:abfit}
  b(a, n) = \left(cn^2 + dn + e \right)a^{f}  + gn + h \,,
\end{equation}
where $n$ is the lower charge state of the CSR (i.e., $n = 1$ for ++/+).
Equation~\ref{eq:abfit} was fit to the data in figure~\ref{fig:model-params}(c) and is shown as a dashed line.
The fit parameters are $c = 0.0036(3)$, $d = -0.025(3)$, $e = 0.11(2)$, $f = 0.21(3)$, $g = -0.012(2)$ and $h = 0.04(2)$, where the parentheses indicates the uncertainty in the least significant digit (i.e. $c = 0.0036 \pm 0.0003$).
The fit is better for elements with higher $a(n=1)$, making this expression less accurate for group IIa and f-block metals.
A physical interpretation of the constants in equation \ref{eq:abfit} is beyond the scope of this work.
However, equations \ref{eq:FCSR} and \ref{eq:abfit} show that $F$ can be expressed solely in terms of the CSR, $a = F(\CSR \rightarrow \infty)$, and $n$:
\begin{equation} \label{eq:FCSR-long}
  F \approx \frac{ a \left( \CSR^{0.3} + 0.256 \right)}{\CSR^{0.3} +  \left( 0.0036 n^2 - 0.025 n + 0.11 \right)a^{0.21} - 0.012 n + 0.296}\ \   \mathrm{V/nm} \\.
\end{equation}
Though interesting, we expect that researchers will find equation \ref{eq:FCSR} more useful than equation \ref{eq:FCSR-long} for calculating $F$ from CSRs in APT data due to its simplicity, and its greater accuracy for ++/+ CSRs and group IIa and f-block elements.

\subsection{Application to APT data} \label{section:results-application}

\begin{figure}[h!]
  \centering
  \includegraphics[width=3.46in]{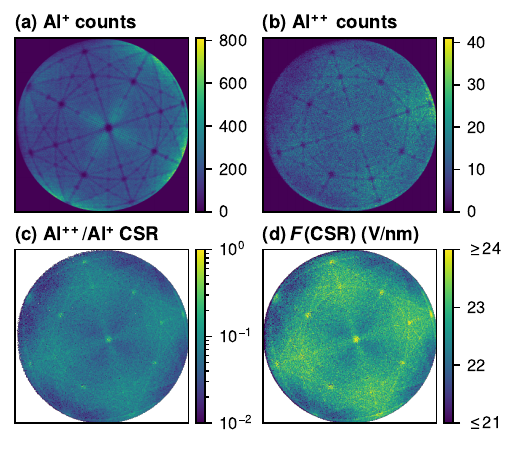}
  \caption{(a) \ce{Al+} and (b) \ce{Al++} field evaporation images from a voltage-pulsed APT experiment of pure Al. (c) The ++/+ CSR for these field evaporation maps. (d) The $F(\CSR)$ calculated using the CSR in~(c), equation \ref{eq:FCSR}, and the Al ++/+ parameters in table \ref{table:all-parameters}.}
  \label{fig:Alhitmap}
\end{figure}

Figure \ref{fig:Alhitmap} shows application of the model $F(\CSR)$ curves to an atom probe dataset of pure Al collected with voltage-pulsed acquisition on a Cameca Invizo 6000.
The reconstruction was performed in detector-space co-ordinates.
The $\left[ 001 \right]$ pole is near the centre of the detector, and three $\left< 111 \right>$ poles are visible at the detector periphery.
The \ce{Al+} and \ce{Al++} peaks were ranged separately and binned into 0.25~nm $\times$ 0.25~nm pixels, producing the field evaporation images shown in figure \ref{fig:Alhitmap} (a,b).
The ratio of these two images is the ++/+ CSR and is shown in (c).
Equation \ref{eq:FCSR} with the model parameters from table \ref{table:all-parameters} produced the $F(\CSR)$ map shown in (d).
The average CSR for this dataset is $5.64 \times 10^{-2}$, corresponding to $F(\CSR) = 22.7$~V/nm.
Figure \ref{fig:Alhitmap}(d) shows the $F$ varies between $\approx 21$~V/nm and $\approx 24$~V/nm across the detector field-of-view, which is $\approx 10\%$ variation around the average value.
As expected \autocite{LarsonSpatial_2013}, the electric field is greater around crystallographic poles.
We have demonstrated that electric fields can be easily calculated from experimental data, and represented in a highly visual way that is simple to interpret.

\begin{figure}[t!]
  \centering
  \includegraphics[width=3.46in]{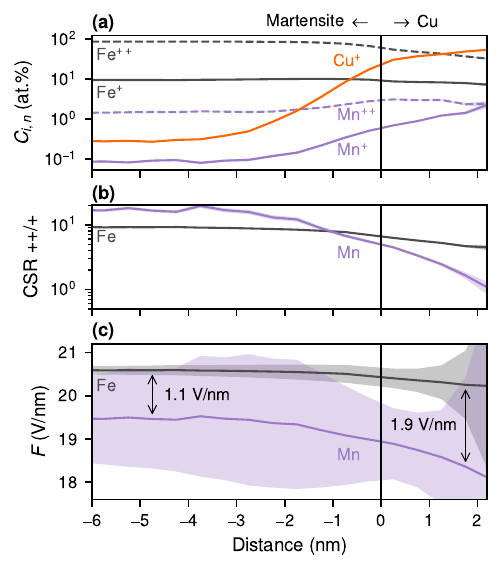}
  \caption{(a) Concentration ($C_{i,n}$) of selected elements ($i$) and charge states ($n$) from a proximity histogram (proxigram) over an iso-concentration surface (isosurface) of Cu ions at 15 at.\% in a microalloyed steel \autocite{Lin_2020}. The + and ++ charge states were observed only for Fe and Mn. Positive distances are inside the Cu precipitates, negative distances in the martensitic matrix. (b) The ++/+ CSR for Fe and Mn. (c) The electric field calculated using equation \ref{eq:FCSR} and the CSRs from (b). The shaded ribbons indicate uncertainties, all propagated from the measurement uncertainty in $C_{i,n}$. The arrows indicate the difference between the $F$ for Fe and Mn in the matrix and the precipitate.}
  \label{fig:Feproxigram}
\end{figure}

Figure \ref{fig:Feproxigram} shows the result of applying the method described here to Cu precipitates in a microalloyed martensitic steel \autocite{Lin_2020}.
An iso-concentration surface (isosurface) of 15 at.\% Cu was calculated on a grid of 0.5 nm spacing with 2 nm of delocalisation.
A proximity histogram (proxigram) of \ce{Cu}, \ce{Fe} and \ce{Mn} is shown in (a).
Only Fe and Mn were observed in both + and ++ charge states, and their CSRs are shown in (b).
The electric field was calculated using equation \ref{eq:FCSR} and table \ref{table:all-parameters} and is shown in (c).
As a relatively low-field metal \autocite{Tsong_1978}, the Cu precipitates have a lower evaporation field than the martensite matrix.
This is reflected in the $\CSR$ and $F$ curves for both metals, and their values drop in the precipitate.
However, the fields predicted by Fe and Mn differ by 1.1~V/nm in the matrix and up to 1.9~V/nm in the precipitate.
This difference is within the range of uncertainty in the $F$ curve for Mn, but this ultimately results from uncertainty in the proxigram and the low concentration of \ce{Mn} in the matrix.
Larger datasets would reduce the uncertainty in $F$, but it's likely there would still be a discrepancy between the $F$ calculated from the Fe and Mn CSRs.

\section{Discussion}

\begin{figure}[h!]
  \centering
  \includegraphics[width=3.46in]{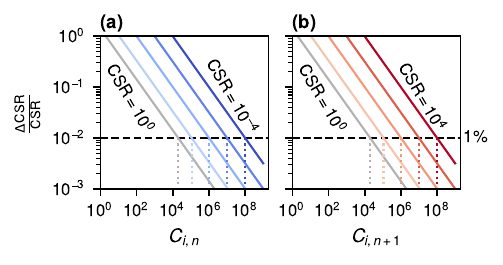}
  \caption{Relative uncertainty in CSR ($\Delta \CSR / \CSR$) in terms of (a) $C_{i,n}$ for $\CSR \leq 10^{0}$ and (b) $C_{i,n+1}$ for $\CSR \geq 10^{0}$. The dashed horizontal line indicates a 1\% relative uncertainty, and the dashed vertical lines indicate the counts required to ensure 1\% relative uncertainty.}
  \label{fig:CSR-relerr}
\end{figure}

The results shown here have been calculated over $10^{-4} \leq \CSR \leq 10^4$, as this approximates the maximum range which can reasonably be measured in a typical APT experiment.
To justify this, we will consider the relative uncertainty across CSRs and dataset sizes.
As weak peaks in mass spectra can be described by a Poisson distribution \autocite{Larson_2013}, the uncertainty in the measured counts are, at worst,
\begin{equation}
  \Delta C_{i,n} = \sqrt{C_{i, n}} \\,
\end{equation}
and similar for the $C_{i, n+1}$ case.
Considering equations \ref{eq:CSRdefinition} and \ref{eq:FCSR}, and assuming $\Delta C_{i,n}$ is not strongly correlated with $\Delta C_{i, n+1}$, the relative uncertainty in CSR is
\begin{equation}
  \frac{\Delta \CSR}{\CSR} \approx \sqrt{\frac{1}{C_{i,n}\CSR } + \frac{1}{C_{i,n}}} = \sqrt{\frac{1}{C_{i,n+1}} + \frac{\CSR}{C_{i,n+1}}}
\end{equation}
where the middle term describes variation with $C_{i,n}$, and
the right term when when considering $C_{i,n+1}$.
Figure \ref{fig:CSR-relerr} shows the relative uncertainty in $\CSR$ for (a) $10^{-4} \leq \CSR \leq 10^0$ with $C_{i,n}$ and (b) $10^{0} \leq \CSR \leq 10^{4}$ with $C_{i,n+1}$.
Each line indicates a $\CSR = 10^k$ for integer $-4 \leq k \leq 4$ and text labels indicate the limits of $k$ for that plot.
The dashed horizontal line indicates the arbitrary uncertainty threshold used in this section, $\Delta \CSR / \CSR = 1\%$.
The dashed vertical lines indicate the counts needed in order to have $1\%$ uncertainty in $\CSR$.
To measure $\CSR = 10^{0}$ at 1\% uncertainty, $C_{i,n} = C_{i,n+1} \approx 10^{4}$ counts are needed in both peaks.
To measure $\CSR = 10^{-4}$ at 1\% uncertainty, $C_{i,n} \approx 10^{8}$ counts are needed in the $n$ peak and thus $C_{i,n+1} \approx 10^{4}$ in the $n+1$ peak.
In general, $\geq 10^{4}$ counts are needed in the smaller peak in order to measure any $\CSR$ with better than $1\%$ accuracy.
Consistently recording datasets of $\gg 100$~million ions without fracture is not routine, even for the newest generation of atom probe instruments \autocite{Tegg_2023}, so accurate measurement of $\CSR \geq 10^{4}$ or $\leq 10^{-4}$ will not be performed frequently.
Additionally, figure \ref{fig:Feproxigram} illustrates how modest uncertainties in $\CSR$ propagate to large uncertainties in $F$.
This is due to the low gradient $\mathrm{d}F / \mathrm{d} \mathrm{CSR}$, particularly for ++/+ CSRs.
Thus when measuring $F$, it's important to minimise uncertainty in $\CSR$ by maximising the sample size used in a calculation.

\begin{figure}[t!]
  \centering
  \includegraphics[width=3.46in]{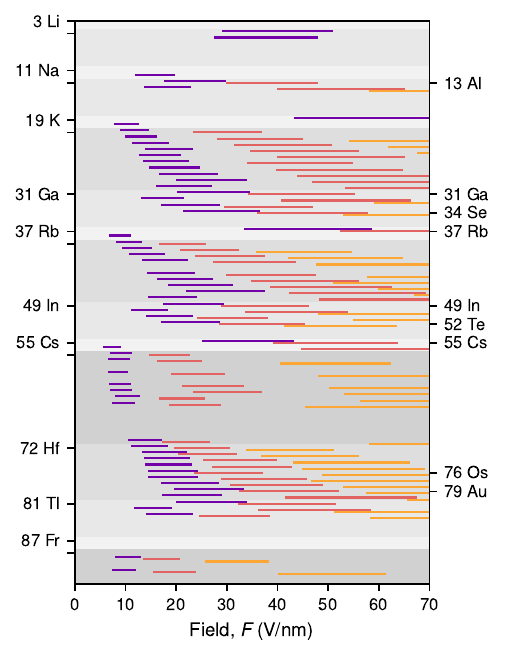}
  \caption{Ranges of $F$ measurable for each element across $10^{-4} \leq \CSR \leq 10^{4}$ using the methods described here. Purple lines indicate ++/+, orange lines +++/++, and yellow lines ++++/+++. The left vertical axis ticks indicate blocks of the periodic table, the right vertical axis indicates elements with overlapping and ranges in $F$.}
  \label{fig:Franges}
\end{figure}

Like Kingham curves, the $F(\CSR)$ curves here can be used to estimate the range of electric fields measurable using CSRs with a given element.
Figure \ref{fig:Franges} shows the ranges in $F$ accessible for each $10^{-4} \leq \CSR \leq 10^{4}$ and element studied in this work.
As with figure \ref{fig:model-params}, elements are sorted by atomic number and separated into blocks.
The right vertical axis shows elements with overlapping ranges where $F(\CSR)$ will be continuously measurable over a wide range of $F$.
For example, it should be possible to measure $F$ using Al ++/+ and +++/++ between $\approx 18 \leq F \leq 45$~V/nm, providing the evaporation rate be managed.
Most of the elements with wide ranges for $F$ measurements are s- or p-block metals, with the addition of \ce{Os} and \ce{Au}.
Similarly, most elements have ranges in $F$ where only a single charge state is expected, and $F$ cannot be measured using $\CSR$ in these ranges. 
An example is Zn, which is expected to be detected only as \ce{Zn++} across $\approx 18 \leq F \leq 52$~V/nm.
In analysis of alloys with multiple matrix elements, it may be possible to use a solute element to measure $F$ if the principle component does not show multiple charge states at the given $F$.
Note that the $F$ range of each CSR is separate to the evaporation field of the element.
For example, the \ce{Zr+} charge state is not commonly observed since the evaporation field of Zr (28~V/nm) \autocite{Tsong_1978} far exceeds the range of its $F(\CSR_{++/+})$ curve.

These calculations were performed only up to ++++.
Although higher charge states have been reported \autocite{Lam_Needs_1992}, the very high fields required to produce these states are not commonly encountered in modern laser-pulsed atom probes \autocite{Tegg_2023} and so were not considered in these calculations.

The model in \autocite{Kingham_1982} is constructed by assuming the tunnelling electron was always leaving an s-type orbital.
Although this is true for the fourth and fifth period d-block metals up to the ++ charge state (e.g., \ce{Ti0} is [Ar] $4\mathrm{s}^2 3\mathrm{d}^1$), the outermost electrons of p-block metals are in p-type orbitals up to ++ (e.g., \ce{Si^0} is [Ne] $3\mathrm{s}^2 3\mathrm{p}^2$), and may be d-type orbitals for the sixth period d-block metals (e.g., \ce{Hf^0} is [Xe] $4\mathrm{f}^{14} 6\mathrm{s}^2 5\mathrm{d}^2)$ \autocite{Mann_pblock_2000, Mann_dblock_2000}.
The low ionisation energies of the f-block metals may also challenge the accuracy of the model, as these were not considered in \autocite{Kingham_1982} or any subsequent literature on post-ionisation.
There is value in repeating the \ce{Rh+ \rightarrow Rh++} field evaporation experiment from \autocite{Ernst_1979} on other metals, as equation \ref{eq:Z} underpins \autocite{Kingham_1982} and all subsequent studies into post-ionisation, including this one.

Section \ref{section:results-application} highlighted some applications to the method described in this work: variation in electric field around poles, and reduction in field within low-field precipitates.
The importance of the electric field to atom probe means there are many other potential applications.
Variations in electric field across precipitates or microstructural features leads to local magnification \autocite{LarsonSpatial_2013}, and knowledge of the electric field allows more sophisticated reconstruction methods \autocite{Lawitzki_2021, Fletcher_2022}.
The electric field affects the relative populations of \ce{^1H+} and \ce{^1H2+} seen in mass spectra \autocite{Mouton_2019}, and controlling the field may allow for discrimination between \ce{^1H2+} and \ce{^2H+} = \ce{D+} peaks at 2~Da.
Acqusition parameters such as laser pulse energy affect the specimen temperature, which in turn affects the electric field needed to induce field evaporation \autocite{Larson_2013}. 
This work aids researchers in thoughtfully choosing acquisition parameters to avoid data issues related to electric field, such as peak overlap or evaporation of metal-hydride species.

\section{Conclusion}

In this work we provide a numerical method for calculating Kingham curves: plots of the probability $P_n$ of observing a charge state $n$ in terms of the electric field $F$ in atom probe tomography (APT) experiments.
Using these data we plot $F$ in terms of the charge state ratio (CSR), and find that simple 2- or 3-parameter expressions can describe $F(\CSR)$ across the first three CSRs and 8 orders-of-magnitude in CSR.
We fit this equation to almost all solid elements on the periodic table and provide a table of constants that allows researchers to calculate $F$ using CSR in their reconstructed APT data.
We illustrate this application using a field evaporation map of pure Al, where we find the evaporation field increases by $\approx 10\%$ from the average value around crystallographic poles.
We also show how the method can estimate the evaporation field inside low-field precipitates in a microalloyed martensitic steel, though calculated values for $F$ differ for the matrix and solute elements.

In general we find our models are most accurate for the ++/+ and +++/++ CSRs, and the p and d-block metals.
Aside from common APT considerations such as maximising mass resolving power and consistently ranging mass spectrum peaks, accurate measurement of $\CSR$ and $F$ requires relatively large datasets.
An uncertainty in CSR of $\leq 1\%$ is only obtainable when there are $\geq 10^{4}$ counts in the smaller of the two peaks used in the CSR calculation.
The simplicity of the models reported here allow researchers to more easily estimate the $F$, allowing for more sophisticated interpretation of phenomena in experimental APT datasets.

\section*{Acknowledgements}

L. Tegg acknowledges S. Huang of the University of Sydney, Australia, for having the scientific problem which inspired this work.
The authors acknowledge Lan (Lance) Yao and Sha (Esther) Li, formerly of the University of Sydney, Australia, for composing an early version of the code used to calculate Kingham curves.
The authors also acknowledge the technical and scientific support provided by Sydney Microscopy and Microanalysis at the University of Sydney, and the support of Microscopy Australia.
The authors acknowledge Takanori Sato of the University of Sydney for providing the pure Al data, and Hung-Wei (Homer) Yen of National Taiwan University for providing the microalloyed steel.

\section*{Supporting information}
There are five supplementary files to this manuscript:
\begin{itemize}
  \item \texttt{CSR.py}: The Python 3.11 script used to perform the calculations in this work. It is divided into sections. Section \texttt{S1} is importing libraries, setting script options, and importing the table of constants. 
  Section \texttt{S2} contains the main calculation loop for $P_n$ and 
  $F(\CSR)$, as well as curve-fitting to $F$, $a$ and $b$. 
  Section \texttt{S3} is for producing tables and plots for inclusion in this manuscript. 
  Section \texttt{S4} is for calculating $F$ from experimental data, and producing plots for inclusion in this manuscript.
  \item \texttt{CSR\_constants.csv}: Atomic and material constants for each element considered in this work \autocite{NIST_ASD, Michaelson_1977}.
  \item \texttt{R6001\_70985\_Al+.csv}: Field evaporation image of \ce{Al+} hits from a specimen of pure Al.
  \item \texttt{R6001\_70985\_Al++.csv}: Field evaporation image of \ce{Al++} hits from a specimen of pure Al.
  \item \texttt{R6001\_40744\_Cu-proxigram.csv}: A proxigram over Cu at 15 at.\% from a microalloyed martensitic steel. This data has not previously been reported, but the material has been described previously in \autocite{Lin_2020}.
\end{itemize}

\printbibliography[heading=bibintoc,title=References]

\end{document}